\documentclass[aps,superscriptaddress,twocolumn,twoside,floatfix,prl,a4paper]{revtex4-2} 
\usepackage{times}
\usepackage{epsfig}
\usepackage{amsfonts}
\usepackage{amsmath}
\usepackage{amssymb,amsthm}
\usepackage{xcolor,colortbl}
\usepackage{multirow}
\usepackage{braket}
\usepackage{latexsym}
\usepackage{amsfonts}
\usepackage{mathrsfs}
\usepackage{natbib}
\usepackage{verbatim}
\usepackage{gensymb}
\usepackage{caption}
\usepackage{caption}
\usepackage{ragged2e}
\DeclareCaptionJustification{justified}{\justifying}
\usepackage{blkarray}
 \usepackage{graphicx}
 \usepackage{enumerate}
\usepackage[shortlabels]{enumitem}
\usepackage{ulem}

\newtheorem{proposition}{Proposition}

\usepackage[colorlinks=true,linkcolor=blue,citecolor=magenta,urlcolor=blue]{hyperref}
\allowdisplaybreaks

\def\>{\rangle}
\def\<{\langle}

\usepackage{centernot}
\usepackage{subfig}

\begin{document}
\title{Randomness-free Test of Non-classicality: a Proof of Concept}

\author{Zhonghua Ma}
\affiliation{Australian Research Council Centre of Excellence for Engineered Quantum Systems \& School of Mathematics and Physics, University of Queensland, QLD 4072, Australia.}

\author{Markus Rambach}
\affiliation{Australian Research Council Centre of Excellence for Engineered Quantum Systems \& School of Mathematics and Physics, University of Queensland, QLD 4072, Australia.}

\author{Kaumudibikash Goswami}
\email{goswami.kaumudibikash@gmail.com}
\affiliation{Australian Research Council Centre of Excellence for Engineered Quantum Systems \& School of Mathematics and Physics, University of Queensland, QLD 4072, Australia.}
\affiliation{QICI Quantum Information and Computation Initiative, Department of Computer Science, University of Hong Kong}

\author{Some Sankar Bhattacharya}
\email{somesankar@gmail.com}
\affiliation{International Centre for Theory of Quantum Technologies, University of Gdansk, Wita Stwosza 63, 80-308 Gdansk, Poland}

\author{Manik Banik}
\affiliation{Department of Physics of Complex Systems, S.N. Bose National Center for Basic Sciences, Block JD, Sector III, Salt Lake, Kolkata 700106, India.}

\author{Jacquiline Romero}
\email{m.romero@uq.edu.au}
\affiliation{Australian Research Council Centre of Excellence for Engineered Quantum Systems \& School of Mathematics and Physics, University of Queensland, QLD 4072, Australia.}

\begin{abstract}
{Quantum correlations and non-projective measurements underlie a plethora of information-theoretic tasks, otherwise impossible in the classical world. Existing schemes to certify such non-classical resources in a {\it device-independent} manner require seed randomness---which is often costly and vulnerable to loopholes---for choosing the local measurements performed on different parts of a multipartite quantum system. In this letter, we propose and experimentally implement a {\it semi-device independent} certification technique for both quantum correlations and non-projective measurements without seed randomness. Our test is {\it semi-device independent} in the sense that it requires only prior knowledge of the dimension of the parts. We experimentally show a novel quantum advantage in correlated coin tossing by producing specific correlated coins from pairs of photons entangled in their transverse spatial modes. We establish the advantage by showing that the correlated coin obtained from the entangled photons cannot be obtained from two 2-level classical correlated coins. The quantum advantage requires performing qubit trine positive operator-valued measures (POVMs) on each part of the entangled pair, thus also certifying such POVMs in a {\it semi-device independent} manner. This proof of concept firmly establishes a new 
{\it cost-effective} certification technique for both generating non-classical shared randomness and implementing non-classical measurements, which will be important for future multi-party quantum communications.}

\end{abstract}



\maketitle
{\it Introduction.--} Correlations play an integral role in information processing be it classical or quantum. Nature presents us with composite systems consisting of correlations among multiple subsystems, that cannot be explained if the subsystems are separable \cite{Einstein1935,Schrodinger1936,Bell1966,Clauser1969,Aspect1982,Pan1998}. Characterizing such non-classical correlations has been central to quantum theory. Aside from testing Bell inequalities, recent developments in quantum technology provide us with the tools to detect non-classicality of correlations either as a pseudo-telepathy game \cite{Brassard1999,Brunner2014} or in a communication task assisted by those correlations \cite{Pawlowski2009,Pawlowski2010}. Both cases involve randomizing over the choice of inputs (measurement settings in the first case and preparation and measurement in the second case). In this work, we implement a new technique of detecting non-classical correlations, which does not require costly seed randomness for inputs. As a trade-off the experimenter is required to know only an upper bound to the dimension of the subsystems in use, hence the technique is semi-device-independent. Besides foundational interest, this new tool paves the way for a cost-effective characterization of non-classical resources in quantum information and computation. 

We follow an operational approach by considering the task of generating shared randomness between two distant parties. Shared randomness (SR), also known as public/correlated randomness (as opposed to private randomness \cite{Fischer2020}), can be thought of as a joint probability distribution over random variables shared between two distant parties, that cannot be factorized. Mutual information is a well-known quantifier of such correlations and is a bonafide measure for the distant parties agreeing on a string of measurement outcomes, given a common source \cite{Bogdanov2011,Ghazi2018,Maurer1993}. Based on mutual information, shared randomness has been established as a useful resource in a number of tasks: privacy amplification, simultaneous message passing, simulation of noisy classical and quantum channels, secret sharing, simulation of quantum correlations, and Bayesian game theory, to name a few \cite{Bennett1988,Bennett1995,Newman1996,Babai,Ahlswede1993,Toner2003,Aumann1987,Cubitt11,Bennett14,Brunner2013,Roy2016,Banik2019,Canonne2017,Patra2022}. Therefore, generation of shared randomness from some physical system is a question of utmost practical relevance. In an operational theory, SR between two distant parties can be obtained from a bipartite system prepared in some correlated state. In practice, the two parties could each be given a part of a correlated pair of classical or quantum coins which they could use for ``coin-tossing". Each party performs a local operation on their respective parts of the composite system which results in correlated outcomes and hence SR.

Here we demonstrate an experimental quantum advantage in generating SR between two parties. Particularly, we show that a two-qubit system prepared in a maximally entangled state can yield some desired SR, that otherwise is unobtainable from the corresponding classical system---two 2-level correlated classical coins which we call two-2-coin. This in turn establishes a non-classical feature of the two-qubit system which distinguishes it from its classical counterpart. Importantly, in our case, a single measurement---positive operator value measure (POVM)---is performed on each part of the entangled pair. Therefore, unlike Bell tests (see \cite{Hall2010,Barrett2011,Banik2013(1)}), no randomization over the choice of local measurements is required for establishing this non-classicality. In the experiment, we use transverse-mode entangled photon pairs produced via degenerate spontaneous parametric down-conversion as the two-qubit system. The advantage is established through a payoff (different from mutual information) in a game played between two distant parties \cite{Guha2021}. The payoff is upper bounded by a threshold value when the parties share a two-2-coin state, whereas a better payoff can be obtained from a two-qubit singlet state even when the state is noisy. The resulting quantum advantage requires generalized measurements, {\it viz.} POVMs \cite{Kraus1983,Busch1986} on the local parts of the shared entangled state. The advantage cannot be obtained from  local projective measurements, {\it a.k.a.} von Neumann measurements \cite{Peres1995} and subsequent post-processing of the outcome statistics. Payoff exceeding the classical threshold value ensures that the measurements are not projective, and thus establishes a semi-device-independent test of generalized measurement.  

{\it Correlated coin tossing.--} The operational utility of SR can be understood within the framework of resource theory \cite{Chitambar2019}. Sources of two random variables $X$ and $Y$ held by two distant parties, Alice and Bob, will not yield any SR whenever the joint probability distribution is in the product form, {\it i.e.}, $P(X,Y){=}P(X)P(Y)$; here $P(Z){\equiv}\{p(z)~|~p(z)\ge0~\&~\sum_{z\in Z} p(z){=}1\}$ denotes a probability distribution on $Z$. On the other hand, the joint source produces a nonzero amount of shared randomness when the distribution cannot be written as a product. The amount of SR can be quantified by the entropic function called mutual information, $I(X{:}Y){:=}H(X){+}H(Y){-}H(X,Y)$; where $H(Z){:=}-\sum p(z)\log_2p(z)$ denotes the Shannon entropy associated with the source $P(Z)$ \cite{Cover2006}. A source $P(Z)$ can be converted into a different one $P^\prime(Z^\prime)$ by a stochastic map $S^{Z\to Z^\prime}$, represented by a $|Z^\prime|\times|Z|$ matrix having non-negative real elements with the entries in each column adding up to unity \cite{Markov2003}. While constructing the resource theory of SR, the free operations on a bipartite source $P(X,Y)$ are given by the product of stochastic maps applied on the individual parts, {\it i.e.}, instead of a general stochastic matrix of the form $S^{XY\to X^\prime Y^\prime}$ only product of local stochastic matrices $S^{X\to X^\prime}$ and $S^{Y\to Y^\prime}$ are allowed as free. For convenience,  the free operations can be represented as a tensor product, $S^{X\to X^\prime}\otimes S^{Y\to Y^\prime}$ \cite{Guha2021}.

\begin{center}
\begin{figure}[t!]
\centering
\includegraphics[height=3cm,width=8.5cm]{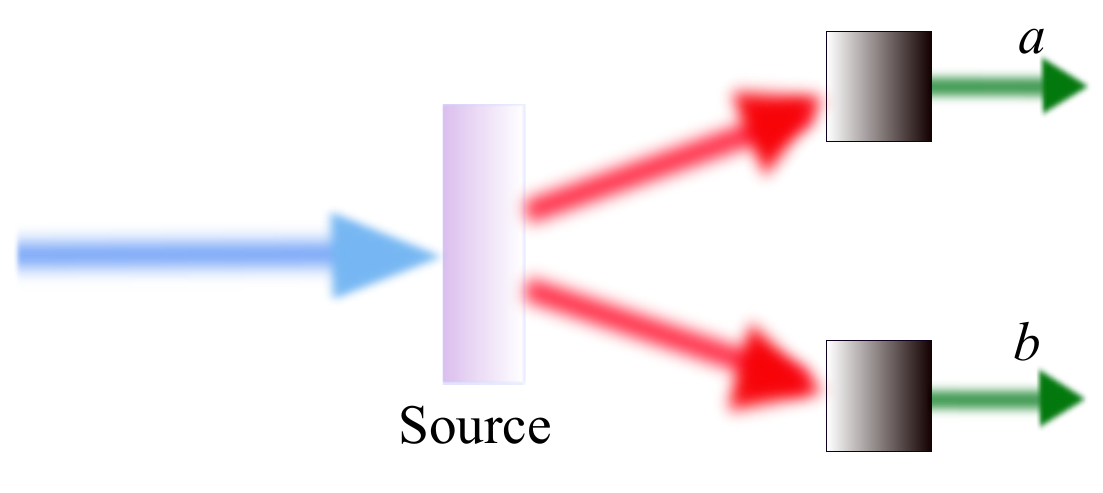}
\caption{A trusted source is emitting bipartite correlated systems of local dimension $2$, which are then being measured by spatially separated black box devices, which outputs $a\in\{1,\dots,d\}$ and $b\in\{1,\dots,d\}$. The sets of observed joint probability distributions $P(a,b)$ for $d>2$, are different for classical and quantum sources.}
\label{fig:my_label}\vspace{-.5cm}
\end{figure}
\end{center}

Physically, SR can be obtained from a composite system prepared in some correlated state which is shared between distant parties. Alice and Bob perform local operations on their respective parts of the composite state resulting in random but correlated outcomes. Within the framework of generalized probability theory, the state space of such a bipartite system is described by $\Omega_{A}\otimes\Omega_{B}$, where $\Omega_{K}$ denotes the marginal state space \cite{Barrett2007}. For instance, the state space of $d$-level classical system is the $d$-simplex, a convex set  embedded in $\mathbb{R}^{d-1}$ having $d$ number of extremal points.
The state space of a two-$d$-coin, shared between two parties Alice and Bob, is defined as, $\mathbf{C}(d)\equiv\{(p(\mathrm{11}),p(\mathrm{12}),\cdots,p(\mathrm{dd}))^{\mathrm{T}}~|~p(\mathrm{ij})\ge0,~\forall~\mathrm{i},\mathrm{j}\in\{\mathrm{1},\cdots,\mathrm{d}\},~\&~\sum_{\mathrm{i},\mathrm{j}}p(\mathrm{ij})=1\}$ (also see \cite{Self0}). The quantum analogue of two-$d$-coin is a two-qudit system associated with the Hilbert space $\mathbb{C}^d_A\otimes\mathbb{C}^d_B$, and the corresponding state space is given by $\mathcal{D}(\mathbb{C}^d_A\otimes\mathbb{C}^d_B)$; where $\mathcal{D}(\mathbb{H})$ denotes the set of density operators acting on the Hilbert space $\mathbb{H}$. From a quantum state, $\rho_{AB}\in\mathcal{D}(\mathbb{C}^d_A\otimes\mathbb{C}^d_B)$, Alice and Bob can generate shared randomness (classically correlated coin) by performing local measurements on their respective parts (see Fig. \ref{fig:my_label}). By $\mathbf{C}(k\to d)$, with $k\le d$, we denote the set of two-$d$-coin states that can be obtained from the set of two-$k$-coin states $\mathbf{C}(k)$ by applying free local stochastic maps $S_{A/B}^{k\to d}$ on Alice's and Bob's part of the states. Similarly, $\mathbf{Q}(k\to d)$ denotes the set of two-d-coin states obtained by performing $d$-outcome local measurements on Alice's and Bob's parts of the bipartite quantum states $\mathcal{D}(\mathbb{C}^k_A\otimes\mathbb{C}^k_B)$. We are now in a position to present our first result as stated in the following proposition (see Appendix A in Supplemental Material \cite{Self} for proof). 
\begin{proposition}\label{prop1}
For every $d\ge3,~\mathbf{C}(2\to d)\subseteq\mathbf{Q}(2\to d)\subsetneq\mathbf{C}(d)$, whereas $\mathbf{Q}(2\to 2)=\mathbf{C}(2)=\mathbf{C}(2\to 2)$. 
\end{proposition} 
As evident from this Proposition, a quantum advantage in SR generation is possible if we consider the generation of a higher dimensional correlated coin state starting from a lower dimensional correlated coin state. More precisely, a proper set inclusion relation $\mathbf{C}(2\to d)\subsetneq\mathbf{Q}(2\to d)$ for some $d\ge 3$ establishes a quantum advantage in correlated coin state generation, which is experimentally testable through the game introduced in \cite{Guha2021}. \vspace{-.8cm}

{\it Quantum advantage in correlated coin tossing.--} Consider a game $\mathbb{G}(n)$ involving Alice and Bob working in an organization. They buy their lunch from one of the $n$ restaurants $\{r_1,\cdots, r_n\}$. The reimbursement policy of the organization ensures the maximum reimbursement of the lunch bill when (1) Alice and Bob do not end up in the same restaurants on the same day, and  (2) their probability of visiting different restaurants should be identical in the long run (see {Appendix B in} \cite{Self} for details). Achieving the maximum reimbursement requires Alice and Bob to share some classical shared randomness which they use to decide which restaurants to go in a particular day.  Satisfying (1) and (2) (perfect success) of the game $\mathbb{G}(3)$ requires Alice and Bob to share a coin $\mathcal{C}_{ac}(3):=\frac{1}{6}\left(0,1,1,1,0,1,1,1,0\right)^{\mathrm{T}}\in\mathbf{C}(3)$. As it turns out, this particular coin cannot be generated from any of the coin states in $\mathbf{C}(2)$ by applying free operations. The optimal payoff Alice and Bob can have with a $\mathbf{C}(2)$ coin is $\mathcal{R}^{\mathbf{C}(2)}_{\max}(3)=1/8<1/6=\mathcal{R}_{\max}(3)$ (see supplemental material). In the quantum case, Alice and Bob, however, can start their protocol by sharing a noisy singlet state 
\begin{align}
\mathcal{W}_p:=p\ket{\psi^-}\bra{\psi^-}+(1-p)\frac{\mathbf{I}_2}{2}\otimes\frac{\mathbf{I}_2}{2},~~p\in[0,1]; \label{werner}   
\end{align}
where $\ket{\psi^-}:=\frac{1}{\sqrt{2}}(\ket{0}\otimes\ket{1}-\ket{1}\otimes\ket{0})$ with $\{\ket{0},\ket{1}\}$ denoting the eigenstates of Pauli $\sigma_z$ operator. Both of them perform the {\it trine} POVM 
\begin{align}
\mathcal{M}\equiv\{e_i&:=\frac{1}{3}(\mathbf{I}_2+\hat{n}_i.\sigma)\}:~~\hat{n}_i:=(\sin\theta_i,0,\cos\theta_i)^{\mathrm{T}},\nonumber\\
\mbox{where}~~&\theta_1=0,~\theta_2=2\pi/3,~\theta_3=4\pi/3,\label{trine}
\end{align}
on their respective qubits. This results in a shared coin state $
\mathcal{C}_p(3):=\left(f_p,s_p,s_p,s_p,f_p,s_p,s_p,s_p,f_p\right)^{\mathrm{T}}\in\mathbf{C}(3)$, with $f_p:=(1-p)/9,~s_p:=(2+p)/18$. This manifests in the payoff 
\begin{align}
\mathcal{R}_p(3):=\min_{i\neq j}P(ij)=(2+p)/18, \label{payoff} 
\end{align}
if Alice and Bob visit the $i^{th}$ restaurant when the $i^{th}$ outcome clicks in their respective measurements. A quantum advantage is demonstrated whenever $\mathcal{R}_p(3)>1/8$. Quantum states $\mathcal{W}_p\in\mathcal{D}(\mathbb{C}^2\otimes\mathbb{C}^2)$ become advantageous over the classical two-2-coin states $\mathbf{C}(2)$ in playing the $\mathbb{G}(3)$ game whenever $p>1/4$, with maximally entangled states yielding the highest payoff \cite{Guha2021}. 

{\it Randomness-free test of non-classicality--} A POVM represents the most general quantum measurement. A POVM is a collection of positive semidefinite operators $\{e_i\}_{i=1}^k$, with $\sum_ie_i=\mathbf{I}_d$, where $\mathbf{I}_d$ is the identity operator acting on the Hilbert space $\mathbb{C}^d$ associated with the system \cite{Kraus1983,Busch1995}. Projective measurements are special cases, where the coefficients $e_i$ correspond to mutually orthogonal projectors $\pi_i$. For a qubit, such a measurement can have only two outcomes: $\{\pi_i:=\ket{\psi_i}\bra{\psi_i}~|~\langle{\psi_i}|\psi_j\rangle=\delta_{ij}~\mbox{for}~~i,j=0,1\}$. A $k$-outcome POVM $\{e_i\}_{i=1}^k$ will be called projective simulable if the outcome probabilities of the POVM elements can be obtained by coarse-graining the outcome probabilities of some $d$-outcome projective measurement for any $d<k$, {\it i.e.} $\forall~i\in[1,k],~e_i=\sum_{j} P_{ij}\pi_{j}$, with $\{\pi_j\}_{j\in[1,d]}$ being a $d$-outcome projective measurement and $\{P_{ij}\}_i$ denoting probability distributions. For instance, the unsharp qubit measurement $\sigma_{\hat{n}}(\lambda)\equiv\{\frac{1}{2}(\mathbf{I}_2\pm\lambda \hat{n}.\sigma)~|~\lambda\in(0,1)\}$ can be simulated through the projective measurement $\sigma_{\hat{n}}\equiv\{\frac{1}{2}(\mathbf{I}_2\pm\hat{n}.\sigma)\}$, since $\frac{1}{2}(\mathbf{I}_2\pm\lambda \hat{n}.\sigma)=\frac{1+\lambda}{4}(\mathbf{I}_2\pm\hat{n}.\sigma)+\frac{1-\lambda}{4}(\mathbf{I}_2\mp\hat{n}.\sigma)$ \cite{Busch1986,Banik2013}. Not all POVMs are projective simulable and such measurements are known to be useful for a number of information-theoretic tasks \cite{Ivanovic87,Peres88,Acin16,Vertesi10,DiMario}. Our game $\mathbb{G}(3)$ provides a semi-device independent certification of such qubit measurements. Denoting the set of all qubit projective simulable measurements as $\mathbf{PS}(2)$, the result is formally stated as the following proposition. \begin{proposition}\label{prop2}
The maximum payoff $\mathcal{R}_{max}^{\mathbf{PS}(2)}(3)$ of the game $\mathbb{G}(3)$, achievable when the players are restricted to perform measurement from the set $\mathbf{PS}(2)$, is upper bounded by $\mathcal{R}_{max}^{\mathbf{C}(2)}(3)$.
\end{proposition}
The claim of Proposition \ref{prop2} follows from the fact that given dimension $d$ of the local sub-systems, the joint outcome probabilities obtained from any arbitrary quantum state and projective measurement are the diagonal elements of the density matrix (the state) when written in the same basis as the projective measurement. Thus the same statistics can also be obtained from a classically correlated (diagonal) state and measurement on the computational basis. 

A payoff higher than the maximum classical payoff, therefore, certifies that the qubit measurements performed by the players are not projective simulable \cite{SSB}. We highlight that this certification technique is semi-device-independent, with the experimenter requiring only the knowledge of the local dimension (in this case $d=2$) of the state shared between parties. As shown in \cite{Gomez16}, certification of POVM is possible even in a device-independent manner. However, such a device-independent certification  requires violation of a suitably designed Bell-type inequality and hence requires each of the parties involved in the Bell test to randomly perform incompatible measurements on their part of the shared system. Note that the technique of \cite{Gomez16} is a detection of non-projective measurement only if subsystem dimension $d=2$, which is further guaranteed by a CHSH violation. In contrast, our semi-device-independent scheme requires a single measurement device for each of the parties, getting rid of the seed randomness in inputs to the measurement devices.

{\it Experimental results.--} The quantum coin used in the experiment is a pair of photons entangled in their transverse spatial modes produced via degenerate spontaneous parametric downconversion (SPDC) \cite{MairOAM}.
We show our experimental schematic in Fig.~\ref{fig:setup} (see Appendix E in \cite{Self}). We implement each element $e_i$ of the trine POVM of Eq.(\ref{trine}) by programming an appropriate hologram onto the spatial light modulators (SLMs E-Series, Meadowlark Optics ). The photons reflected from the SLM are coupled to single-mode fibres via lenses L5 and L6. The single-mode fibres are then connected to superconducting nanowire single-photon detectors (SNSPDs, Opus One, Quantum Opus). The probability of a party going to a particular restaurant is proportional to the probability of the outcomes of the POVM which can then be obtained from photon counts. To get the joint probability, we record the coincidence count ($C_{ij}$) between Alice's $i$th and Bob's $j$th measurement outcome.  This is done via a time-tagging module (TT20, Swabian Instruments) by integrating for 3600 seconds per data point. We normalise the coincidence counts to evaluate the joint probability $P(ij)$ for Alice and Bob going to the $i^{th}$ and $j^{th}$ restaurant, i.e., $P(ij)=C_{ij}/\sum_{i,j}C_{ij}$ and evaluate the payoff of Eq.\,(\ref{payoff}).

\begin{figure}[b!]
\includegraphics[height=5cm,width=8.5cm]{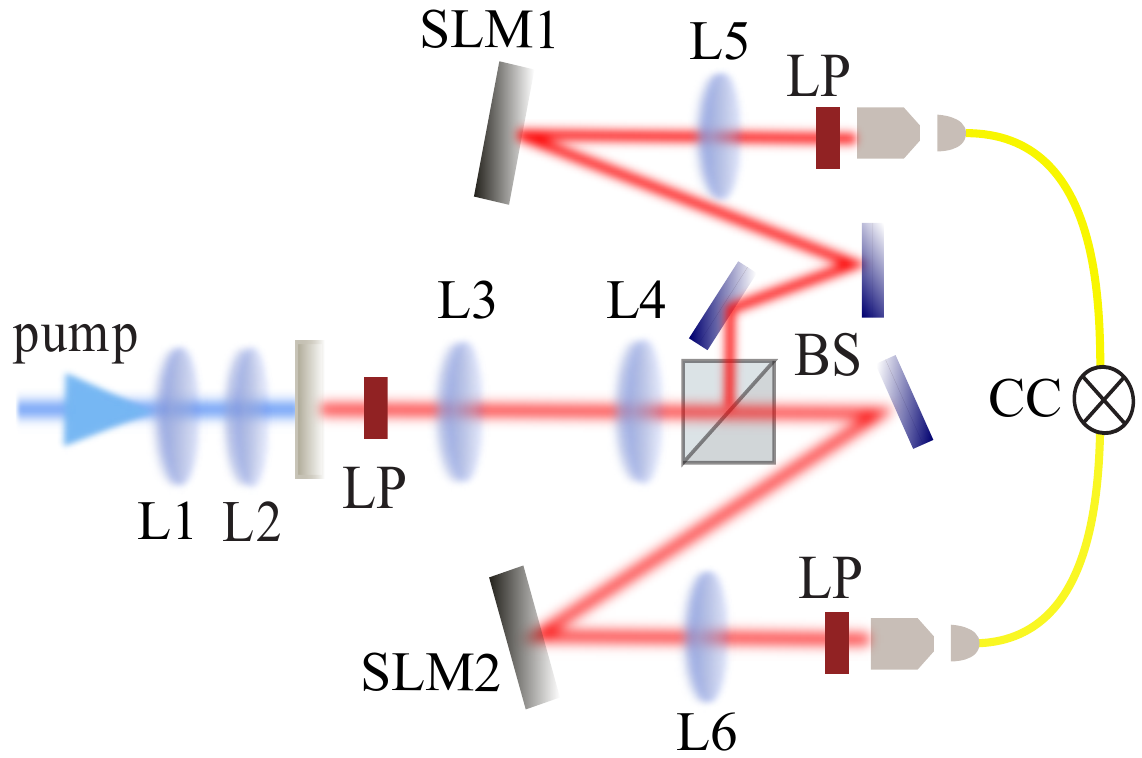}
\caption{Experimental Setup. A 405-nm pump laser (pump) goes through a nonlinear crystal (BBO) producing pairs of entangled photons at 810 nm. A long-pass filter (LP) separates the pump from the single photons, which are then split probabilistically by a 50/50 beamsplitter (BS) between the signal and idler arms representing Alice's and Bob's shares. The combination of the spatial light modulators (SLM1 and SLM2), single mode fibres (yellow curves), and single photon detectors (not shown)  correspond to measurements of the transverse mode of the single photons. The output of the single photon detectors is fed to a coincidence circuit (CC) to record the correlations. Lenses (Ls) are placed along the optical path to optimise mode matching.}
\label{fig:setup}
\end{figure}

The entangled photons produced by SPDC are in the state $\ket{\psi^+}$, which transforms to the $\ket{\psi^-}$ state when a $\sigma_z$-rotation is applied to one of the photons. Alternatively, we program the hologram for measuring $\sigma_z e_i \sigma_z$ for one of the photons, where $e_i$ is a POVM element as defined in Eq.(\ref{trine}). In the same manner, for the noisy case where the quantum coin is in the noisy state of Eq.(\ref{werner}), instead of generating $\mathcal{W}_p$, we add the noise to the measurements. The state $\mathcal{W}_p$ signifies that one of the subsystems of the singlet state $\ket{\psi^-}$ undergoes a depolarising channel of strength $p$, {\it i.e.}, the state remains unchanged with a probability $(1{+}3p)/4$ or undergoes any of the three Pauli rotations, each with a probability $(1{-}p)/4$. In our experiment, we introduce this depolarising channel in the measurement settings by implementing the POVM-element $e_i$ affected by the noise. This can be done by measuring $\{e_i\}$ with a probability of $(1{+}3p)/4$ and measuring $\{\sigma_j e_i \sigma_j\}$ with a probability of $(1{-}p)/4$, where $\{\sigma_j\}$ represent the Pauli operations. Experimentally we implement this noisy POVM by performing a weighted time average on the relevant measurements. For a total acquisition time of $T$, we measure $\{e_i\}$ for a time duration of $T(1{+}3p)/4$ and measure $\{\sigma_j e_i \sigma_j\}$ for a time duration of $T(1{-}p)/4$ each. Thus the temporal degree of freedom is used as pointers for the Kraus-operators of the depolarising channel, and time-averaging erases this pointer information leading to a statistical mixture of the Kraus-operators \cite{Shaham_controllable_depol}.

\begin{figure}[t!]
\includegraphics[width=\columnwidth]{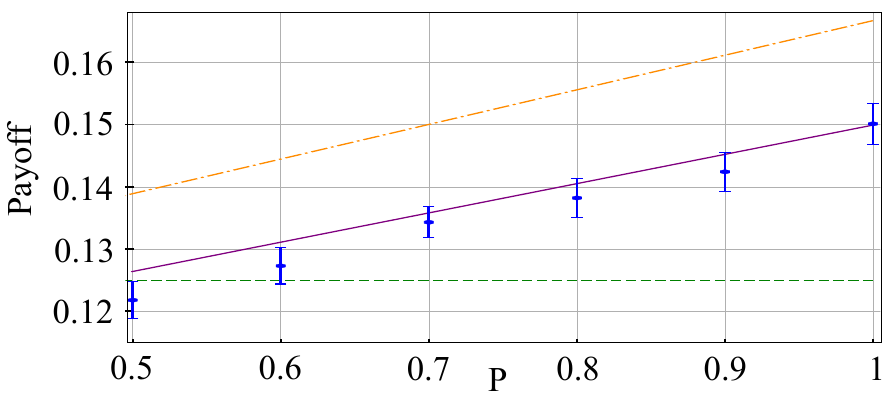}
\vspace{-.6cm}
\caption{Payoff of classical optimum strategy vs. quantum strategy. 
The optimal classical payoff of $0.125$ is shown as the dashed green line. Ideal quantum payoff following Eq.(\ref{payoff}) is plotted in a dash-dotted orange line. Theoretically expected payoffs considering the imperfect entangled state are shown in the solid purple line. Payoffs obtained in experiments are shown in blue dots (along with error bars) and they are all above the classical limit for $p>0.6$.}
\label{fig:payoff_exp}
\end{figure}

Results obtained in the experiment are depicted in Fig. \ref{fig:payoff_exp}. 
The payoffs from the probabilities obtained from our experiment are all above the classical limit of 0.125 (dashed green line) for $p>0.6$, with the highest value being $0.150\pm0.003$ obtained for the noiseless case.  The experimental payoffs as a function of the noise (denoted by the depolarisation strength $p$) are given by the blue dots.  The ideal payoffs are given by the dash-dotted orange line (i.e. if we have a perfect maximally entangled state for the correlated coins and perfect POVMs). The discrepancies between the experiment and the ideal case can be accounted for by our imperfect entangled state. The purple solid line is the expected payoff given the entangled state that we obtained via quantum state tomography using an over-complete set of 36 measurements (97.0\% fidelity to $\ket{\psi^+}$ and purity of 95.2\%). Nevertheless, our experiment firmly establishes a quantum advantage in correlated coin-tossing even with a significant amount of noise. Apart from establishing the advantage in generating shared randomness our experiment also has another interesting implication as it constitutes a semi-device independent certification of non-projective measurement.

{\it Discussions.--} 
Tests of the quantum nature of physical systems are complicated by the requirement of randomness in the inputs to such tests. For example, the true randomness that quantum systems are known to exhibit can only be certified in a device-independent manner using Bell's theorem \cite{Pironio2010}, which in turn needs true randomness (however small) in the inputs \cite{Putz2014}-- at least qualitatively the argument is circular. The quantum advantage for shared randomness processing that we experimentally demonstrate in this paper is important as it provides a way to test non-classical correlations without the need for true randomness---the only test of this kind. Our method certifying both non-classical correlations and generalised measurements is semi-device-independent, requiring only knowledge of the dimensionality of the subsystem.   We show that a two-qubit system prepared in a maximally entangled state leads to a higher payoff in our two-party game by yielding some desired correlated coin state, which is  impossible to obtain from any two 2-level correlated classical coins. 
In contrast to the advantage in randomness processing demonstrated in \cite{Dale2015,Patel2019} which involves the probability distribution of one random variable, our work focuses on a new kind of quantum advantage in generating {\it shared} randomness between two distant parties, involving two random variables and their joint probability distributions.  This latter quantum advantage will find use in distributed computational tasks, as in the example of the game we show here. Our work sets the stage for further studies of quantum advantage in multiparty shared randomness processing for qubits or even higher-dimensional systems. Given that randomness processing is an important computational primitive, we envision our work will be useful for information processing in quantum networks. 

\begin{acknowledgements}
We thank Valerio Scarani for helpful discussions. SSB acknowledges support from the Foundation for Polish Science (IRAP project, ICTQT, contract no. MAB/2018/5, co-financed by EU within Smart Growth Operational Programme). This research was supported by the Australian Research Council Centre of Excellence for Engineered Quantum Systems (EQUS, CE170100009) and Discovery Project (DP200102273). MB acknowledges funding from the National Mission in Interdisciplinary Cyber-Physical systems from the Department of Science and Technology through the I-HUB Quantum Technology Foundation (Grant no: I-HUB/PDF/2021-22/008),  support through the research grant of INSPIRE Faculty fellowship from the Department of Science and Technology, Government of India, and the start-up research grant from SERB, Department of Science and Technology (Grant no: SRG/2021/000267). JR is supported by a Westpac Bicentennial Foundation Research Fellowship.
\end{acknowledgements}

\onecolumngrid

\section{Supplemental Material}

\section{Appendix A: Proof of Proposition 1}
\begin{proof}
First, note that both $\mathbf{C}(2\to d)$ and $\mathbf{Q}(2\to d)$ consist of valid probability vectors in $\mathbf{C}(d)$, by definition. Second, to show that $\mathbf{C}(2\to d)\subseteq \mathbf{Q}(2\to d)$, we need to show that for every choice of classically correlated coin states in $\mathbf{C}^2$ and stochastic map $S_{A}^{2\to d}$ and $S_{B}^{2\to d}$ for Alice and Bob respectively, there exist two-qubit states in $\mathbf{Q}(2)$ and $d$ outcome qubit POVMs such that the final joint probability of Alice's and Bob's outcomes are the same in both cases. 

Let us denote an arbitrary state in $\mathbf{C}(2)$ as a column vector
\begin{align}\label{eq:p1}
\mathcal{C}(2)&=(a_{00},a_{01},a_{10},a_{11})^{\mathrm{T}},\\
a_{ij}\ge 0,&~\forall~i,j\in \{0,1\},~\&~\sum_{i,j} a_{ij}=1,\nonumber
\end{align}
and stochastic matrices 
\begin{align}\label{eq:p2}
S_{A}^{2\to d}=(P_{k,i})_{\substack{k= 0,\dots,d-1 \\ i=0,1}},~~~ S_{B}^{2\to d}=(Q_{l,j})_{\substack{l= 0,\dots,d-1 \\ j=0,1}},\\ 
P_{k,i},Q_{l,j}\ge0,~~\sum_kP_{ki}=1=\sum_lQ_{lj}.\nonumber  
 \end{align} 
 The resulting coin state is given by,
 \begin{align}
 \mathcal{C}'(d)=S_A^{2\to d} \otimes S_B^{2\to d} [\mathcal{C}(2)].
\end{align}
The proof is by construction: the two-d-coin state $\mathcal{C}'(d)$ can also be obtained from the operators 
 \begin{align}
\rho=\sum_{i,j=0,1}a_{ij}|ij\rangle\langle ij|,~~~~~~~~~~~~~~~\\   e_{A}^k=\sum_{i}P_{k,i}|i\rangle\langle i|,~~~   e_{B}^l=\sum_{j}P_{l,j}|j\rangle\langle j|,\\
\mbox{such that,}~~ \mathcal{C}'(d)=(Tr[\rho (e_{A}^k\otimes e_{B}^l)])_{k,l}.
 \end{align}
 Now it only remains to show that $\rho$ is a valid two qubit state and $\{e_{A}^k\}_{k=0}^d-1$ and $\{e_{B}^l\}_{k=0}^d-1$ are valid POVMs. First, we observe that Eq. (\ref{eq:p1}) implies that $\rho\ge0$ (positive semi-definite), and $Tr[\rho]=\sum_{i,j} a_{ij}=1$ (unit trace). From Eq. (\ref{eq:p2}) we have, $e_{A}^k\ge0$, $e_{B}^l\ge0$ for all $k,l=\{0,\dots,d-1\}$ (positivity) and $\sum_{k}e_{A}^k=\sum_{k}\sum_{i}P_{k,i}|i\rangle\langle i|=\sum_{i}(\sum_{k}P_{k,i})|i\rangle\langle i|=\sum_{i}|i\rangle\langle i|=\mathbb{I}$ (normalization). Similarly, $\sum_{l}e_{B}^l=\sum_{l}\sum_{j}P_{l,j}|j\rangle\langle j|=\sum_{j}(\sum_{l}P_{l,j})|j\rangle\langle j|=\sum_{j}|j\rangle\langle j|=\mathbb{I}$. Hence, this construction leads to the fact that all $d$-outcome joint outcome probabilities obtained by Alice and Bob, when accessing a classical source of shared randomness of local dimension $2$, can also be obtained with a two-qubit state and $d$-outcome POVMs.      
\end{proof}

\section{Appendix B: Non-monopolizing social subsidy game (Dining out in pandemic times)} 
The game $\mathbb{G}(n)$ involves two employees, Alice and Bob, working in an organization. There are $n$ different restaurants $\{r_1,\cdots, r_n\}$ where the employees can choose to buy their daily lunch. The organization follows a reimbursement policy to pay back the bill. For this, each day’s bills of Alice and Bob are collected for a log to calculate the joint probability $P(ij)$ that Alice visits restaurant $r_i$  while Bob visits restaurant $r_j$ restaurant. Alice and Bob are reimbursed only if they end up going to different restaurants, {\it i.e.}, $r_i\neq r_j$. At the same time, the game requires minimising the chance of one restaurant being boycotted, which would happen if Alice and Bob frequent two different but fixed restaurants. Hence we define the payoff as, 
\begin{align}
\$\mathcal{R}(n) =\$\min_{i\neq j} P(ij)
\end{align}
to ensure that trade is distributed among all the restaurants. The employees are non-communicating. However, they possess a bipartite state with subsystems described by two-level systems and choose strategies with the help of free operations and accordingly try to maximize their payoff. If no restriction is put on the amount of SR, then both Alice and Bob can obtain the maximum payoff \cite{Guha2021} 
\begin{align}
\mathcal{R}_{\max}(n)=\frac{1}{(n^2-n)}~.   
\end{align}
The situation becomes interesting if restrictions are imposed on the amount of SR. Let, $\mathcal{R}^{\mathbf{C}(m)}_{\max}(n)$ denote the maximum payoff achieved in $\mathbb{G}(n)$ when the parties are allowed to share a two-$m$-coin. For a pair $(m,n)$ with $m<n$, the cases $\mathcal{R}^{\mathbf{C}(m)}_{\max}(n)<\mathcal{R}_{\max}(n)$ open up a scope for quantum advantage.

\section{Appendix C: Playing the game $\mathbb{G}(n)$ with a $\mathbf{C}(d)$ coin}\label{AppenB}
We will first show that the game $\mathbb{G}(3)$ can not be perfectly won if Alice and Bob share any of the two-2-coin states from the set $\mathbf{C}(2)$. As already discussed in the main manuscript, the perfect winning strategy in $\mathbb{G}(3)$ game demands Alice and Bob to share the coin state $\mathcal{C}_{ac}(3)\in\mathbf{C}(3)$. Our aim is to show that the coin state $\mathcal{C}_{ac}(3)$ does not belong to the set $\Theta\mathrm{C}(2\to 3)$, a subset of $\mathbf{C}(3)$ that can be obtained from the set of two-2-coin states $\mathbf{C}(2)$ by applying free local stochastic maps $S_{A/B}^{2\to 3}$ on Alice's and Bob's part of the states. Let they start with an arbitrary state in $\mathbf{C}(2)$: 
\begin{align}
\mathcal{C}(2)&=(a_{00},a_{01},a_{10},a_{11})^{\mathrm{T}},\\
a_{ij}\ge 0,&~\forall~i,j\in \{0,1\},~\&~\sum_{i,j} a_{ij}=1.\nonumber
\end{align}
Alice and Bob apply the following local stochastic maps on their respective part of the coin state,
\begin{align}
S_{A}^{2\to 3}=
  \left[ {\begin{array}{ccccc}
    P_{00} & P_{01}\\
    P_{10} & P_{11}\\
    P_{20} & P_{21}
  \end{array} } \right],~~~ S_{B}^{2\to 3}=
  \left[ {\begin{array}{ccccc}
    Q_{00} & Q_{01}\\
    Q_{10} & Q_{11}\\
    Q_{20} & Q_{21}\\
  \end{array} } \right];\\ 
P_{ij},Q_{ij}\ge0,~~\sum_iP_{ij}=1=\sum_iQ_{ij}.~~~~~~~~~~\nonumber
\end{align}
The resulting coin state is given by,
\begin{align}\label{eq:c}
\mathcal{C}'(3)= S_A^{2\to 3} \otimes S_B^{2\to 3} [\mathcal{C}(2)].  
\end{align}
Comparing the state in Eq. (\ref{eq:c}) with $\mathcal{C}_{ac}(3):=\frac{1}{6}\left(0,1,1,1,0,1,1,1,0\right)^{\mathrm{T}}$, we obtain the following equations
\footnotesize
\begin{subequations}
\begin{align}
    a_{00} P_{00} Q_{00} + a_{10} P_{01} Q_{00} + a_{01} P_{00} Q_{01} + a_{11} P_{01} Q_{01} &= 0,\label{eq:c1}\\
  a_{00} P_{10} Q_{10} + a_{10} P_{11} Q_{10} + a_{01} P_{10} Q_{11} + a_{11} P_{11} Q_{11} &= 0, \label{eq:c2}\\
  a_{00} P_{20} Q_{20} + 
  a_{10} P_{21} Q_{20} + 
  a_{01} P_{20} Q_{21} + a_{11} P_{21} Q_{21} &= 0, \label{eq:c3}\\
  a_{00} P_{00} Q_{10} + a_{10} P_{01} Q_{10} + a_{01} P_{00} Q_{11} + a_{11} P_{01} Q_{11} &= 1/6, \label{eq:c4}\\
  a_{00} P_{00} Q_{20} + a_{10} P_{01} Q_{20} + a_{01} P_{00} Q_{21} + a_{11} P_{01} Q_{21} &= 1/6, \label{eq:c5}\\
  a_{00} P_{10} Q_{00} + a_{10} P_{11} Q_{00} + a_{01} P_{10} Q_{01} + a_{11} P_{11} Q_{01} &= 1/6, \label{eq:c6}\\
  a_{00} P_{10} Q_{20} + a_{10} P_{11} Q_{20} + a_{01} P_{10} Q_{21} + a_{11} P_{11} Q_{21} &= 1/6,\label{eq:c7}\\
  a_{00} P_{20} Q_{00} + a_{10} P_{21} Q_{00} + a_{01} P_{20} Q_{01} + a_{11} P_{21} Q_{01} &= 1/6, \label{eq:c8}\\
  a_{00} P_{20} Q_{10} + a_{10} P_{21} Q_{10} + a_{01} P_{20} Q_{11} + a_{11} P_{21} Q_{11} &= 1/6. \label{eq:c9}
\end{align}
\end{subequations}
\normalsize
Now the following cases are possible:
\begin{itemize}
\item[]{\bf Case (A):} Let us assume $a_{ij}\neq 0$ for all $i,j$. From Eq. (\ref{eq:c1}) we must have the following minimum sets of parameters with {\it zero} value:
\begin{align*}
{\bf (A0)}: \{P_{00},P_{01}\}=0,~~~~
{\bf (A1)}: \{Q_{00},Q_{01}\}=0.
\end{align*}
Condition {(\bfseries A0)} contradicts Eq. (\ref{eq:c4}) since it implies that the right-hand side of Eq. (\ref{eq:c4}) is zero. Similarly, conditions {(\bfseries A1)} contradicts Eq. (\ref{eq:c6}).
\item[]{\bf Case (B):} Let us assume $a_{00}=0$ and $a_{ij}\neq 0$ for $ij\neq 00$. From Eq. (\ref{eq:c1}) we must have the following:
\begin{align*}
{\bf (B0)}: \{P_{00},P_{01}\}=0,~~~~
{\bf (B1)}: \{P_{01},Q_{01}\}=0.
\end{align*}
Same as the previous case, Condition {(\bfseries B0)} contradicts Eq. (\ref{eq:c4}). From Eq. (\ref{eq:c8}), Condition {(\bfseries B1)} implies $P_{21}=P_{11}=\frac{1}{2}$. Now Eq. (\ref{eq:c2}) along with this condition implies 
\begin{align*}
{\bf (B2)}: \{Q_{10},Q_{11}\}=0.   
\end{align*}
Finally, condition {(\bfseries B2)} contradicts Eq. (\ref{eq:c4}).
\item[]{\bf Case (C):} Let us assume $a_{00}=a_{01}=0$ and $a_{10},a_{11}\neq 0$. From Eq. (\ref{eq:c1}) and (\ref{eq:c2}) we must have the following:
\begin{align*}
{\bf (C0)}: \{P_{01},P_{11}\}=0,~~~~
{\bf (C1)}: \{P_{01},Q_{10},Q_{11}\}=0. 
\end{align*}
Both Conditions {(\bfseries C0)} and {(\bfseries C1)} contradicts Eq. (\ref{eq:c4}). 
\item[]{\bf Case (D):} Let us assume $a_{00}=a_{01}=a_{10}=0$ and $a_{11}\neq 0$. From Eq. (\ref{eq:c1}) and (\ref{eq:c2}) we must have the following:
\begin{align*}
{\bf (D0)}: \{P_{01},P_{11}\}=0,~~~~
{\bf (D1)}: \{Q_{01},Q_{11}\}=0,\\
{\bf (D2)}: \{P_{01},Q_{11}\}=0,~~~~
{\bf (D3)}: \{P_{11},Q_{01}\}=0. 
\end{align*}
Both Conditions {(\bfseries D0)} and {(\bfseries D1)} contradicts Eq. (\ref{eq:c4}). Condition {(\bfseries D2)} contradicts (\ref{eq:c4}). Similarly, Condition {(\bfseries D3)} contradicts (\ref{eq:c6}). 
\end{itemize}
Thus in every case, we find a contradiction to the set of Equations (\ref{eq:c1}-\ref{eq:c9}), leading to the fact that the game $\mathbb{G}(3)$ can not be perfectly won if Alice and Bob share any of the two-2-coin state from the set $\mathbf{C}(2)$. 

A natural question is given a two-2-coin what will be the maximum success probability achievable in the game $\mathbb{G}(3)$. As shown in \cite{Guha2021}, this boils down to solving the following optimization problem.  Given
a two-2-coin, $\mathcal{C}(2)\equiv (p(11), p(12), p(21 , p(22))^{\mathrm{T}}$, Alice and Bob can obtain a two-3-coin states $\mathcal{C}(3)\equiv q(11), q(12),q(13), q(21 , q(22),q(23),q(31),q(32),q(33))^{\mathrm{T}}$ by applying local stochastic maps $\mathcal{S}_A^{2\to3}$ and $\mathcal{S}_B^{2\to3}$ on their respective part of the two-2-coin, and aim to maximize the payoff in $\mathbb{G}(3)$ game, {\it i.e.},
\begin{align}
\mathcal{R}_{\max}^{\mathbf{C}(2)}(3)&=\mbox{maximize}~~q(\mathrm{i}\neq \mathrm{j})\nonumber\\
&~~~~~~{\mathcal{C}(2)\in\mathbf{C}(2)}\nonumber\\
&~~~~~~\mathcal{S}_A^{2\to3}\otimes\mathcal{S}_B^{2\to3}\nonumber\\
&~~~~~\mbox{subject~to~}~q(\mathrm{i}\neq \mathrm{j})\le q(\mathrm{i}^\prime\neq \mathrm{j}^\prime)\nonumber\\
&~~~~~~~\mathrm{i}\neq \mathrm{i}^\prime~\mbox{and/or}~\mathrm{j}\neq \mathrm{j}^\prime~.\label{opt}
\end{align}
As it turns out, the maximum payoff Alice and Bob can achieve is $1/8$. One possible optimal classical strategy is the following: Alice and Bob start with a shared coin $\mathcal{C}_{\alpha}(2):=\left(\alpha,0,0,\alpha\right)^{\mathrm{T}}\in\mathbf{C}(2)$, with $\alpha=1/2$; and apply local stochastic maps $S^{2\to3}_A=\begin{pmatrix}0&\alpha&\alpha\\\alpha&0&\alpha
\end{pmatrix}^{\mathrm{T}}
$ and $S^{2\to3}_B=\begin{pmatrix}\alpha&0&\alpha\\0&\alpha&\alpha
\end{pmatrix}^{\mathrm{T}}$ on their respective parts. This results in the following $\mathbf{C}(3)$ coin 
\begin{align*}
\left[\begin{pmatrix}0&\frac{1}{2}\\\frac{1}{2}&0\\\frac{1}{2}&\frac{1}{2}\end{pmatrix}\bigotimes \begin{pmatrix}\frac{1}{2}&0\\0&\frac{1}{2}\\\frac{1}{2}&\frac{1}{2}
\end{pmatrix}\right]\begin{pmatrix}\frac{1}{2}\\0\\0\\\frac{1}{2}\end{pmatrix}=\frac{1}{8}\begin{pmatrix}0\\1\\1\\1\\0\\1\\1\\1\\2\end{pmatrix},
\end{align*}
which assures the payoff $\mathcal{R}^{\mathbf{C}(2)}_{\max}(3)=1/8$.

We have established a quantum advantage in shared randomness processing, by showing that the game $\mathbb{G}(3)$ cannot be perfectly won by any two-faced correlated classical coin $\mathbb{C}(2)$, but Alice and Bob have the perfect strategy if they are allowed to share a two-qubit state. However, the quantum advantage over the classical resource is limited in this case -- the two-faced correlated classical coin $\mathcal{C}_{ac}(3)\in\mathbb{C}(3)$ yields the perfect strategy to play the $\mathbb{G}(3)$ game. Interestingly, the advantage can be made stronger. For that, consider the game $\mathbb{G}(4)$ which, for the perfect success $\mathcal{R}_{\max}(4)=1/12$, requires the coin state 
\begin{align}
\mathcal{C}_{ac}(4)&=(p(11),p(12),p(13),p(14),p(21),p(22),\nonumber\\
&~~~~~~~~p(23),p(24),p(31),p(32),p(33),p(34),\nonumber\\
&~~~~~~~~p(41),p(42),p(43),p(44))^{\mathrm{T}}\nonumber\\
&=\left(0,\frac{1}{12},\frac{1}{12},\frac{1}{12},\frac{1}{12},0,\frac{1}{12},\frac{1}{12},\frac{1}{12},\frac{1}{12},\right.\nonumber\\
&~~~~~~~~~~~\left.0,\frac{1}{12},\frac{1}{12},\frac{1}{12},\frac{1}{12},0\right)^{\mathrm{T}}\in\mathbf{C}(4).
\end{align}
The optimal success probability can be achieved by sharing the two-qubit singlet state and by performing SIC-POVM instead of trine-POVM in our experiment.  Following similar optimization as in Eq.(\ref{opt}) it can be shown that $\mathcal{R}^{\mathbf{C}(2)}_{\max}(4)=1/15$ and $\mathcal{R}^{\mathbf{C}(3)}_{\max}(4)=2/27$. Thus, in the $\mathbb{G}(4)$ game a two-qubit system exhibits an advantage over the $3$-faced correlated classical coins. But, the quantum gain (difference between optimal quantum success and optimal classical success) is reduced. While in the $\mathbb{G}(3)$ game the quantum gain is $(1/6-1/8)=1/24$, in the $\mathbb{G}(4)$ game the quantum gain over the $\mathbf{C}(2)$ coins is $(1/12-1/15)=1/60$ and over the $\mathbf{C}(3)$ coins it is $(1/12-2/27)=1/108$. From an experimental point of view,  a lower quantum gain is not desirable as it limits noise tolerance in the experiment.

\section{Appendix D: Device-independent status of Quantum advantage in correlated coin tossing}
In view of an adversarial scenario, one may consider a third party Eve, and ask: is it possible for Eve to obtain information about $P(X,Y)$, the joint probability distribution of the random variables $X$ and $Y$, which are outcomes of Alice's and Bob's measurements? The answer to such a question definitely depends on the computational power of Eve and the constraints on the system shared between Alice and Bob. For the case where Alice and Bob achieve perfect success in the $\mathbb{G}(3)$ game, it is not possible for Eve to obtain information about $P(X,Y)$, provided that Eve's power is restricted to process quantum systems only and Alice's and Bob's local subsystems' dimensions are bounded from above (in our case dim$=2$). This follows from the observation that perfect success necessitates a maximally entangled two-qubit state to be shared between Alice and Bob, which cannot be correlated even classically with a third party. However, a general answer to the question of security in this setting for arbitrary success probability $\mathcal{R}_p(3)>1/8$, we leave as a question for future research.

\section{Appendix E: Experimental setup}
We describe the experimental setup in this section. A 405-nm continuous wave ultraviolet laser is incident on a 5-mm thick beta-barium borate (BBO) crystal (after passing through a pair of lenses of focal length 50 mm to control collimation, L1 and L2, see Fig.2 in the main body of the paper). While passing through the crystal, a UV photon is probabilistically downconverted to a pair of infrared photons of wavelength 810 nm. We block the remaining UV photons by a long-pass filter (LP) and separately image the infrared photons to two spatial light modulators (SLM1 and SLM2) by the combination of lenses L3 (f=150 mm), L4 (f=300 mm) and the beam splitter (BS).  The SPDC process conserves optical orbital angular momentum (OAM), denoted by $\ell$ which can take on any integer value, and hence is theoretically infinite-dimensional (in practice, the apertures in the experiment limit the dimensionality). We use the SLMs to project the entangled state onto the qubit space with $\ell = \pm3$, i.e. the Bell state $\ket{\psi^+}$ with $\ket{0} \equiv \ket{\ell=-3}$ and $\ket{1} \equiv \ket{\ell=3}$. Choosing a value for $|\ell|$ is a trade-off between the high extinction of the two qubit states and high coincidence count rates. As the number of measurements is limited to nine settings in our experiment, we prioritised higher extinction, while compensating for lower rates with longer integration times. The amplitude of the individual contributions of the two-qubit state is balanced via angle tuning of the BBO crystal~\cite{Romero2012}. We obtained a coincidence rate of 120 (143) per 60 seconds for the $\ket{3,-3}$ ($\ket{-3,3}$) state measurements, with a coincidence-to-accidental ratio (CAR) of $\sim 43$. The CAR could be further increased past 500 by appropriately narrowing the coincidence time window from 25~ns to around 2~ns, without loss of any signal coincidences from our source.

\subsection{Description of the holograms for the measurement}

To perform measurements on the OAM basis, we need to display the corresponding holograms on the spatial light modulator (SLM). The hologram, in this context, refers to the phase mask that we program on the SLM.  The combination of this hologram and the single mode-fibre is a rank-1 POVM on the state of the single photon that is impinging on the SLM. This method of measuring the transverse spatial mode of a single photon was introduced by Mair et al.\cite{MairOAM} in the first demonstration of the entanglement of the orbital angular momentum of photons.

\section{Appendix F: Generating correlated coin from $\mathbb{C}^d\otimes\mathbb{C}^d$ systems}
Our experiment establishes that a $2$-qubit state prepared as a singlet state provides an advantage over the $2$-faced correlated classical coin states in generating shared randomness. An interesting question is how to generalize this result, {\it i.e.} how to establish the advantage of a $2$-qudit quantum system over the $d$-faced correlated classical coin states in shared randomness processing. In the absence of proof of the persistence of a quantum advantage even for higher dimensions, we propose a candidate correlation obtained from $2$-qudit maximally entangled state, that may not be achievable with comparable classical coins. 

Our experiment involves the $2$-qubit singlet state 
\begin{align*}
\ket{\psi^-_2}_{AB}:=\frac{1}{\sqrt{2}}\left(\ket{0}_A\otimes\ket{1}_B-\ket{1}_A\otimes\ket{0}_B\right)\in\mathbb{C}^2\otimes\mathbb{C}^2.   
\end{align*}
The $2$-qubit singlet has an interesting property in that it yields anti-correlated outcomes when Alice and Bob perform the identical spin measurements in any possible direction, {\it i.e.},   
\begin{align}
\ket{\psi^-_2}_{AB}=\frac{1}{\sqrt{2}}\left(\ket{\chi_0}_A\otimes\ket{\chi_1}_B-\ket{\chi_1}_A\otimes\ket{\chi_0}_B\right),\label{ac}
\end{align}
for any orthonormal basis $\{\ket{\chi_0},\ket{\chi_1}\}$ of $\mathbb{C}^2$. this particular property plays a crucial role in establishing the desired advantage. 

Consider another $2$-qubit maximally entangled state
\begin{align*}
\ket{\phi^+_2}_{AB}:=\frac{1}{\sqrt{2}}\left(\ket{0}_A\otimes\ket{0}_B+\ket{1}_A\otimes\ket{1}_B\right)\in\mathbb{C}^2\otimes\mathbb{C}^2.   
\end{align*}
This state also has a different symmetry, {\it i.e.}
\begin{align}
\ket{\phi^+_2}_{AB}=\frac{1}{\sqrt{2}}\left(\ket{\chi_0}_A\otimes\ket{\chi_0^\star}_B+\ket{\chi_1}_A\otimes\ket{\chi_1^\star}_B\right),\label{c2}    
\end{align}
for any orthonormal basis $\{\ket{\chi_0},\ket{\chi_1}\}$ of $\mathbb{C}^2$, where for a state $\ket{\chi}=\alpha\ket{0}+\beta\ket{1}$ we have $\ket{\chi^\star}=\alpha^\star\ket{0}+\beta^\star\ket{1}$. This symmetry holds true for the state $\ket{\phi^+_d}_{AB}\in\mathbb{C}^d\otimes\mathbb{C}^d$, {\it i.e.}
\begin{align}
\ket{\phi^+_d}_{AB}&=\frac{1}{\sqrt{d}}\sum_{i=0}^{d-1}\ket{i}_A\otimes\ket{i}_B\nonumber\\
&=\frac{1}{\sqrt{d}}\sum_{i=0}^{d-1}\ket{\chi_i}_A\otimes\ket{\chi^\star_i}_B,  
\end{align}
where $\{\ket{\chi_i}\}_{i=0}^{d-1}$ is an orthonormal basis of $\mathbb{C}^d$. 
Now if Alice and Bob perform suitably chosen SIC-POVM (possibly Bob's SIC-POVM could be constructed with each element being conjugate-orthogonal to a corresponding element of Alice's SIC-POVM) on their respective parts of the aforementioned entangled state then they can generate the following interesting coin state, 
\begin{align*}
\mathcal{C}^\star_{ac}(d)=(p(ij))^{\mathrm{T}},~~i,j\in\{1,\cdots,d\}\\
\mbox{where},~~p(i=j)=0~~\&~~p(i\neq j)\neq0.
\end{align*}
Please note that the coin states $\mathcal{C}_{ac}(d)$ (previously discussed perfectly anti-correlated coin states) are also described by the states $\mathcal{C}^\star_{ac}(d)$. In $\mathcal{C}_{ac}(d)$ all the probabilities $p(i\neq j)$'s are identical, while this condition is relaxed in $\mathcal{C}^\star_{ac}(d)$ as it is only demanded that all the probabilities $p(i\neq j)$'s must be non-zero. 

Now, for $d=3$ case, perfect anti-correlation implies $p(ij)=\frac{1}{72}$ for all $i\neq j$ (not known to be realizable with $2$-qutrit states). Analogous to the $d=2$ case, the payoff being the minimum of the probabilities of the anti-correlated events, would be $\frac{1}{72}$. A trivial classically realizable payoff is $\frac{1}{81}$ (for completely local random strategy), which should make the gap between the quantum and classical payoff smaller than $\frac{1}{648}$, which will reduce the noise tolerance of the scheme to a great extent. Another point of complexity to observing the classical vs quantum gap in higher dimensions for the present scheme stems from the fact that the optimization for the classical protocols is a non-convex problem, which involves a polynomially growing number of parameters with increasing dimensions and the number of outcomes. 

The other avenue to establish quantum advantage could be to show that the class of states $\mathcal{C}^\star_{ac}(3)$ are not realizable by $2$-three coin states, similar to the proof in Appendix B\ref{AppenB}. Such an analytical proof becomes quite cumbersome (although not impossible) given the large number of parameters.


\begin{thebibliography}{99}
\bibitem{Self0} A normal coin-tossing experiment corresponds to the case $d{=}2$, and leads to the state space $\mathbf{C}^2{\equiv}\{(p(\mathrm{h}),p(\mathrm{t}))^{\mathrm{T}}~{|}~p(\mathrm{h}),p(\mathrm{t}){\ge}0~\&~p(\mathrm{h}){+}p(\mathrm{t}){=}1\}$; $p(\mathrm{h})$ and $p(\mathrm{t})$ respectively denote the probability of obtaining outcome `head'  and `tail'. The state space $\mathbf{C}(2)\equiv\mathbf{C}^2_A\otimes\mathbf{C}^2_B$ of two-2-coin shared between Alice and Bob is given by $\mathbf{C}(2)\equiv\{(p(\mathrm{hh}),p(\mathrm{ht}),p(\mathrm{th}),p(\mathrm{tt}))^{\mathrm{T}}~|~p(\mathrm{ij})\ge0,~\forall~\mathrm{i},\mathrm{j}\in\{\mathrm{h},\mathrm{t}\},~\&~\sum_{\mathrm{i},\mathrm{j}}p(\mathrm{ij})=1\}$.
\end{thebibliography}
\end{document}